\documentclass[a4paper,11pt]{article}

\usepackage[dvips]{graphicx}

\begin{document}

\begin{center}
{\large \bf Nuclear multifragmentation and fission: similarity and differences}

\vspace{10mm}

    V. Karnaukhov$^{1,}$\footnote{Email address: karna@jinr.ru}, H. Oeschler$^2$, S. Avdeyev$^1$, V. Rodionov$^1$, \\
    V. Kirakosyan$^1$, A. Simonenko$^1$, P. Rukoyatkin$^1$, A. Budzanowski$^3$, \\
   W. Karcz$^3$, I. Skwirczynska$^3$, B. Czech$^3$, L. Chulkov$^4$, E. Kuzmin$^4$, \\
    E. Norbeck$^5$, A. Botvina$^6$

\vspace{5mm}

 { \small
 $^1$Joint Institute for Nuclear Research, 141980 Dubna, Russia\\
 $^2$Institut f\"ur Kernphysik, Darmstadt University of Technology, 64289 Darmstadt, Germany\\
 $^3$H. Niewodniczanski Institute of Nuclear Physics, 31-342 Cracow,Poland \\
 $^4$Kurchatov Institute, 123182 Moscow, Russia\\
 $^5$University of Iowa, Iowa City, IA 52242, USA\\
 $^6$Institute for Nuclear Research, 117312 Moscow, Russia\\}

\end{center}

\vspace{0.5cm}

\noindent Thermal multifragmentation of hot nuclei is interpreted as the nuclear {\it liquid--fog}
phase transition deep inside the spinodal region. The experimental data for p(8.1GeV) + Au collisions
are analyzed. It is concluded that the decay process of hot nuclei is characterized by {\it two size
parameters}: transition state and freeze-out volumes. The similarity between dynamics of
fragmentation and ordinary fission is discussed. The IMF emission time is related to the mean rupture
time at the multi-scission point, which corresponds to the kinetic freeze-out configuration.

\section{\normalsize Introduction}

\hspace{4mm} The study of the highly excited nuclei is one of the challenging topics of nuclear
physics, giving access to the nuclear equation of state for temperatures below ${\it T}_{c}$ -- the
critical temperature for the {\it liquid-gas} phase transition. The main decay mode of hot nuclei is
a copious emission of intermediate mass fragments (IMF), which are heavier than $\alpha$--particles
but lighter than fission fragments. An effective way to produce hot nuclei is via heavy-ion
collisions. But in this case the heating of nuclei is accompanied by compression, strong rotation,
and shape distortion, which influence the decay properties of excited nuclei. One gains simplicity,
and the picture becomes clearer, when light relativistic projectiles (protons, antiprotons, pions)
are used. In this case, fragments are emitted by only one source -- the slowly moving target
spectator. Its excitation energy is almost entirely thermal. Light relativistic projectiles provide a
unique possibility for investigating {\it thermal multifragmentation}. The decay properties of hot
nuclei are well described by statistical models of multifragmentation (SMM and MMMC \cite{1,2}). This
is an indication, that the system is thermally equilibrated or close to that. For the case of
peripheral heavy-ion collisions the partition of the excited system is also governed by heating.

The van der Walls equation can be used with nuclear matter because of the similarity of the
nucleon-nucleon force to the force between molecules in a classical gas \cite{3,4,5}. In both cases
there exists a region in the PVT diagram corresponding to a mixture of liquid and gas phases. This
region can contain unstable, homogeneous matter for short times. In a classical gas this can be
achieved by cooling through the critical point.  In the nuclear case this can be achieved by a sudden
expansion of the liquid phase at a temperature well below the critical temperature. The separation of
the homogeneous matter into a mixture of stable liquid and gas is called spinodal decomposition. One
can imagine that a hot nucleus (at ${\it T}$ = 7--10 MeV) expands due to thermal pressure and enters
the unstable region. Due to density fluctuations, a homogeneous system is converted into a mixed
phase consisting of droplets (IMF) and nuclear gas interspersed between the fragments. Thus the final
state of this transition is a {\it nuclear fog} \cite{5}. Note that classical fog is unstable, it
condensates finally into bulk liquid. The charged nuclear fog is stable in this respect. But it
explodes due to Coulomb repulsion and is detected as multifragmentation. It is more appropriate to
associate the spinodal decomposition with the {\it liquid-fog} phase transition in a nuclear system
rather than with the {\it liquid-gas} transition \cite{6,8}. This scenario is supported by a number
of
observations; some of them are the following:\\
\noindent (a) the density of the system at break-up is much lower than the normal one ${\rho}_{0}$
\cite{8};\\ (b) the mean life-time of the fragmenting system is very small ($\approx$ 50 {\it fm/c})
\cite{9};\\
(c) the break-up temperature is significantly lower than $T_{c}$, the critical temperature for the
{\it liquid-gas} phase transition \cite{6,7}.\\
\indent In this paper we concentrate on the dynamics of thermal multifragmentation and its similarity
to ordinary fission. This similarity was noted first by Lopez and Randrup in their statistical theory
of multifragmentation \cite{10,11}. First of all, there are two characteristic volumes (or
configurations) for both processes. Secondly, the time scale characterizations for fragmentation and
fission are similar with respect to their ingredients.\\
\indent Experimental data have been obtained using the $4\pi$-device FASA installed at the external
beam of the Nuclotron (Dubna). At present, the setup consists of twenty five {\it dE-E} telescopes
surrounded by a fragment multiplicity detector, which is composed of 58 thin CsI(Tl) scintillation
counters.\\

\section{\normalsize Two characteristic volumes in thermal multifragmentation}

\hspace{3mm} Traditionally, in statistical models \cite{1,2}, multifragmentation is characterized by
just one size \\
\noindent parameter -- the freeze-out volume, ${\it V}_{f}$. There are a number of papers with
experimental estimations of this characteristic volume, but the values obtained deviate
significantly. A mean freeze-out volume $\sim$ 7${\it V}_{0}$ (${\it V}_{0}$ = volume at normal
density) was found in $\it ref.$ \cite{12} from the average relative velocities of the IMFs for
$^4$He(14.6 MeV) + Au collisions. In paper \cite{13} the nuclear caloric curves were considered
within the Fermi-gas model to extract average nuclear densities for different systems. It was found
that ${\it V}_{f}$ $\approx$ 2.5${\it V}_{0}$ for the medium and heavy masses. In $\it ref.$
\cite{14} the mean IMF kinetic energies were analyzed for Au + Au collisions at 35$\cdot$A MeV. The
freeze-out volume was found to be $\sim$ 3${\it V}_{0}$.  The average source density for the
fragmentation in the 8.0 GeV/$\it c$ ${\pi}^{-}$ + Au interaction was estimated to be $\approx$
(0.25--0.30)${\rho}_{0}$, at E*/A$\sim$5
MeV from the moving-source-fit Coulomb parameters \cite{15,16}.\\
\indent In our paper \cite{8}, the data on the charge distribution and kinetic energy spectra of IMFs
produced in {\it p}(8.1GeV)+Au collisions were analyzed using the combined INC +Exp+SMM model. The
events with IMF multiplicity {\it M}$\ge$2 were selected. The results obtained are shown at {\it
fig.}1. It was shown that one should use two volume (or density) parameters to describe the process
of multifragmentation, not just one as in the traditional approach. The first, ${V}_{t}$ =
(2.6${\pm}$0.3)${\it V}_{0}$ (or ${\rho}_{t}$ ${\approx}$ 0.38${\rho}_{0}$), corresponds to the stage
of pre-fragment formation. Strong interaction between pre-fragments is still significant at this
stage.
\begin{figure}
\begin{center}
  \includegraphics[width=8.1cm]{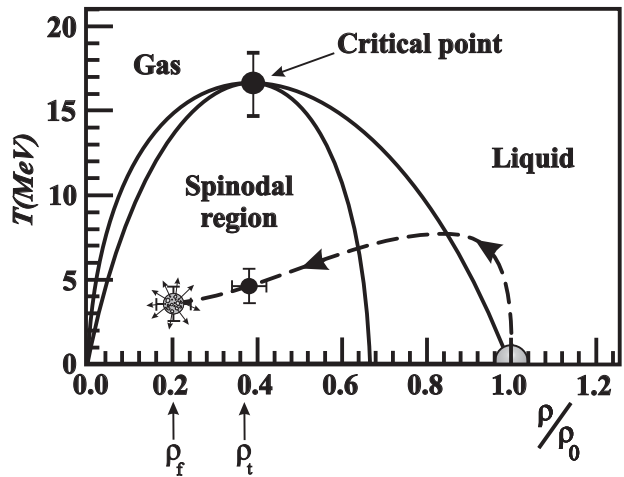}\\
  \caption{Proposed spinodal region for the nuclear system. The experimental points were obtained by
the FASA collaboration. The arrow line shows the way  from the starting point at {\it T}=0 and normal
nuclear density ${\rho}_{0}$ to the break-up point at ${\rho}_{t}$ and to the kinetic freeze-out at
the mean density ${\rho}_{f}$. Critical temperature is estimated in \cite{7}.}\label{1}
\end{center}
\end{figure}
\noindent The second one, ${\it V}_{f}$=(5${\pm}$1)${\it V}_{0}$, is the kinetic freeze-out volume.
It is determined by comparing the measured fragment energy spectra with the model predictions using
multi-body Coulomb trajectories. The calculations have been started with placing all charged
particles of a given decay channel inside the freeze-out volume ${\it V}_{f}$. In this configuration
the fragments are already well separated from each other, they are interacting via the Coulomb force
only. Actually, the system at freeze-out belongs to the mixed phase sector of the phase diagram, with
a mean density that is five times less than normal nuclear density. The first characteristic volume,
${\it V}_{t}$, was obtained by analyzing the IMF charge distributions, {\it Y}(Z), within the SMM
model with a {\it free} size parameter {\it k}. For simplicity, the dependence of the charge
distribution on the
critical temperature ${\it T}_{c}$ was neglected in this analysis.\\
\indent Recently, we performed a more sophisticated consideration of {\it Y}(Z) with two free
parameters, ${V}_{t}$ and ${\it T}_{c}$. A comparison of the data with the calculations was done for
the range 3 $<$ Z $<$ 9, in which minimal systematic errors were expected. {\it Figure }2 shows
${\chi}^{2}$ for comparison of the measured and calculated IMF charge distributions as a function of
${\it V}_{t}$ /${\it V}_{0}$ for different values of the critical temperature. The minimum value of
${\chi}^{2}$ decreases with increasing ${\it T}_{c}$ in the range from 15 to 19 MeV and saturates
after that. The corresponding value of ${\it V}_{t}$/${\it V}_{0}$ increases from 2.4 to 2.9. The
measured IMF charge distribution is well reproduced by SMM with ${\it V}_{t}$ in the range
(2.5--3.0)${\it V}_{0}$, which is close to the value obtained in \cite{8}. As for the kinetic
freeze-out volume, the present value coincides with the one given in \cite{8}, but its uncertainty is
only half as much
because the estimated systematic error is less:\\
\noindent ${V}_{f}$ = (5.0 ${\pm}$0.5)${\it V}_{0}$.\\
\indent Our previous conclusion about the value of the critical temperature \cite{6,7} is also
confirmed: ${\it T}_{c}$ exceeds 15 MeV. Note, this value is twice as large as estimated in [17,18]
with the Fisher droplet model. This contradiction is waiting for further efforts to clarify the point.
\begin{figure}
\begin{center}
\includegraphics[width=7.6cm]{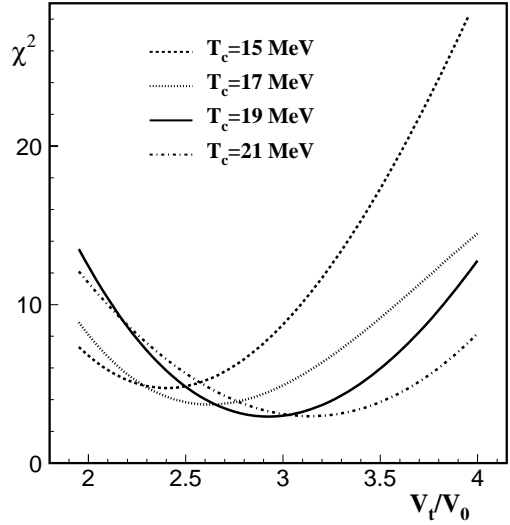}\\
  \caption{Value of ${\chi}^{2}$ as a function of ${\it V}_{t}/{\it V}_{0}$ for comparison of the measured and model predicted IMF charge
distributions. The calculations with the INC*+SMM combined model was performed under the assumption
of two free parameters: ${\it V}_{t}$ -- the effective volume at the stage of pre-fragment formation,
and ${\it T}_{c}$ -- the critical temperature for the liquid-gas phase transition.}\label{2}
\end{center}
\end{figure}
\section{\normalsize Comparison of multifragmentation and fission dynamics}

\hspace{3mm} The occurrence of two characteristic volumes for multifragmentation has a transparent
meaning. The first volume, ${\it V}_{t}$, corresponds to the fragment formation stage at the top of
the fragmentation barrier. Here, the properly extended hot target spectator transforms into closely
packed pre-fragments. The final channel of disintegration is completed during the evolution of the
system up to the moment, when receding and interacting pre-fragments become completely separated at
${\it V}_{f}$. This is just as in ordinary fission. The saddle point (which has a rather compact
shape) resembles the final channel of fission having already a fairly well defined mass asymmetry.
Nuclear interaction between fission pre-fragments ceases after the descent of the system from the top
of the barrier to the scission point. In the papers by Lopez and Randrup \cite{10,11} the similarity
of the two processes was used to develop a theory of multifragmentation based on a generalization of
the transition-state approximation first suggested by Bohr and Wheeler in 1939. The transition states
are located at the top of the barrier or close to it. The phase space properties of the transition
states are decisive for the further fate of the system, for specifying the final channel. No size
parameters are used in the theory.\\
\begin{figure}
\begin{center}
  \includegraphics[width=6.5cm]{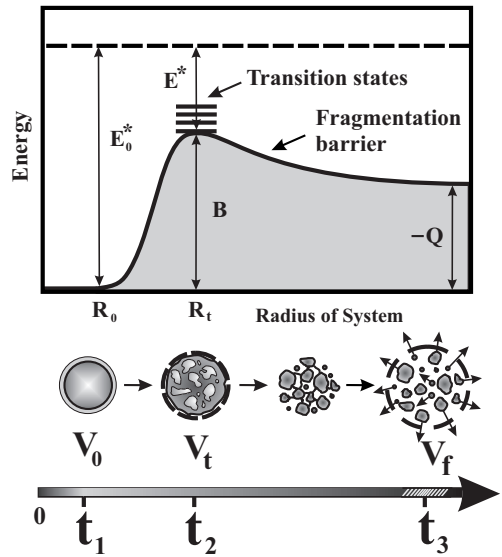}\\
  \caption{Upper: qualitative presentation of the potential energy of the hot nucleus (with excitation energy
${\it E}_{0}^{*}$) as a function of the system radius. The ground state energy of the system
corresponds to {\it E}=0. {\it B} is the fragmentation barrier, {\it Q} is the released energy.
Bottom: schematic view of the multifragmentation process and its time scale: ${\it t}_{1}$ --
thermalization time, ${\it t}_{2}$ -- time of the expansion driven by thermal pressure, ${\it t}_{3}$
-- the mean time of the multi-scission point (with dispersion, which is measured as fragment emission
time, ${\tau}_{em}$ ).}\label{3}
\end{center}
\end{figure}
\indent Being conceptually similar to the approach of {\it ref.} \cite{10,11}, the statistical
multifragmentation model (SMM) uses a size parameter that can be determined by fitting to data. The
size parameter obtained from the IMF charge distribution can hardly be called a freeze-out volume.
This is the "transition state volume", corresponding to the top of the fragmentation barrier (see
{\it fig.}3). The freeze-out volume, ${\it V}_{f}$, corresponds to the multi-scission point, when
fragments became completely separated and start to be accelerated in the common electric field. In
the statistical model (SMM) used, the yield of a given final channel is proportional to the
corresponding statistical weight. This means that the nuclear interaction between pre-fragments is
neglected when the system volume is ${\it V}_{t}$ and that this approach can be viewed as a
simplified transition-state approximation. Nevertheless, the SMM well describes the IMF charge (mass)
distributions for thermally driven multifragmentation. Note once again that in the traditional
application of the SMM, only one size parameter is used. The shortcoming of such a simplification of
the
model is obvious now.\\
\indent The evidence for the existence of two characteristic multifragmentation volumes changes the\\
\noindent understanding the time scale of the process (see the bottom of {\it fig.}3). One can
imagine the following ingredients of the time scale: ${\it t}_{1}$ -- the mean thermalizataion time
of the excited target spectator, ${\it t}_{2}$ -- the mean time of the expansion to reach the
transition state, ${\it t}_{3}$ -- the mean time up to the multi-scission point.\\
\indent The system configuration on the way to the scission point contains several pre--fragments
connected by necks. Their random rupture is characterized by the mean time, ${\tau}_{n}$, which seems
to be a decisive ingredient of the fragment emission time:
 ${\tau}_{em}$ ${\approx}$ ${\tau}_{n}$.
Formally ${\tau}_{em}$ may be understood as the standard deviation of ${t}_{3}$:\\

\hspace{40mm} ${\tau}_{em}$ =$(<{t}^{2}_{3}>-<{t}_{3}>^{2})^{1/2}$.\\

\noindent In the earlier papers, the emission time was related to the mean time of density
fluctuations in the system at the stage of fragment formation, at {\it t}
${\approx}$ $t_{2}$ \cite{19}.\\
\indent What are the expected values of these characteristic times? Thermalization or energy
relaxation time after the intranuclear cascade, ${\it t}_{1}$, is model estimated to be (10--20) {\it
fm/c} \cite{20,21}. The Expanding Emitting Source model (EES) predicts
 $<$${\it t}_{2}$ -- ${\it t}_{1}$$>$ ${\approx}$ 70 {\it fm/c}
for {\it p}(8.1GeV)+Au collisions \cite{22}. The model calculation in \cite{23} results in estimation
of ${\it t}_{3}$ to be (150--200) {\it fm/c}. The only measured temporal characteristic is a fragment
emission time, ${\tau}_{em}$, which is found in number of papers to be ${\approx}$ 50 {\it fm/c} ({\it
e.g.} see
\cite {9}). It would be very important to find a way to measure the value of ${\it t}_{3}$.\\
\indent Note, that in the case of ordinary fission ${\it t}_{2}$ is specified by the fission width
${\Gamma}_{f}$, which corresponds to a mean time of about ${10}^{-19} s$ (or $3.3 {\cdot} {10}^{4}$
{\it fm/c}) for an excitation energy of around 100 MeV [24]. The value $<$${\it t}_{3}$ -- ${\it
t}_{2}$$>$ is model estimated in a number of papers to be about 1000 {\it fm/c} \cite{25}. A mean
neck rupture time, considered as a Rayleigh instability, is estimated in \cite{26} to be:

\vspace{5mm}

\begin{equation}
 {\tau}_{n}=
 [1.5{\cdot}{({R_n}/{\it fm})}^3]^{1/2}{\cdot}{10}^{-22} s
\end{equation}

\vspace{5mm}

\noindent Generally, the values of ${\tau}_{n}$ are found to be 200--300 {\it fm/c} for fission.\\
\indent Using (1) for the estimation of the mean time for the rupture of the multi-neck configuration
in fragmentation, one gets ${\tau}_{n}$ between 40 and 115 {\it fm/c} under the assumption of a neck
radius $R_n$ between 1 and 2 {\it fm}.  This estimation is in qualitative agreement with the measured
values of the fragment emission time ${\tau}_{em}$. Evidently, the multifragmentation process is much
faster than high energy fission. {\it Table} 1 summarizes these
estimates.\\

\vspace{5mm}

\begin{center}
\noindent{\it Table} 1 Characteristic times (in {\it fm/c}) for fragmentation and
fission; the models used are given in brackets; experimental values are marked (Exp).
\end{center}

\begin{center}
\begin{tabular}{|c|c|c|c|c|} \hline
  & ${\it t}_1$ & ${\it t}_2 $ & $<$ ${\it t}_{3}$ -- ${\it t}_{2}$ $>$ &
  ${\sigma}({\it t}_{3})$ \\
\hline
 &  &  &  & \\
Fragmentation & 20   &  80   &  150  & $50{\pm}10$ \\
              & (UU) & (EES) & (QMD) &  (Exp) \\
 &  &  &  & \\
 \hline
 &  &  &  & \\
    Fission   &      &  $3 {\cdot} {10}^{4}$  & $ 2 {\cdot} {10}^{3}$ & 200  \\
              &  & (Exp) & (LD) &  (RI) \\
 &  &  &  & \\

\hline

\end{tabular}
\end{center}

\vspace{10mm}

\indent As for the spatial characteristics, the relative elongation of the very heavy systems (Z$>$99)
at the fission scission point is similar to that for the multi-scission point of medium heavy nuclei.
For the fission of the lighter nuclei, (Po--Ac), the scission elongation is
larger \cite{26}.\\
\indent A few words about the experimental possibility for finding the total time scale for
fragmentation, {\it i.e.} the mean value of ${\it t}_{3}$. It can be done by the analysis of the
fragment-fragment correlation function with respect to relative angle. But, in contrast to the usual
IMF-IMF correlation, one of the detected fragments should be the particle ejected during the
thermalization time ${\it t}_{1}$. For the light relativistic projectiles, it may be the
pre-equilibrium IMF's; for heavy-ion induced multifragmentation, a projectile residual (PR) may be
used as the trigger related to the initial collision. In the last case the PR--TIMF correlation
function should be measured, where TIMF is the intermediate mass fragment from the disintegration of
the hot target spectator created via the partial fusion.

\vspace{4mm}

\section{\bf Conclusion}

\hspace{4mm} Thermal multifragmentation of hot nuclei is interpreted as the nuclear {\it liquid--fog}
phase transition inside the spinodal region. Experimental evidence is presented for the existence of
two characteristic volumes for the process: transition state and kinetic freeze-out volumes. This is
similar to that for ordinary fission. The dynamics is similar also for the two processes, but
multifragmentation is much faster than high energy fission. The IMF emission time is related to the
mean rupture time at the multi-scission point, which corresponds to the freeze-out configuration.\\

\indent The authors are grateful to A. Hrynkiewicz, A.I. Malakhov, A.G. Olchevsky for support and to
I.N. Mishustin and W. Trautmann for illuminating discussions. The research was supported in part by
the Russian Foundation for Basic Research, Grant ¹ 06-02-16068, the Grant of the Polish
Plenipotentiary to JINR, Bundesministerium f\"ur Forschung und Technologie, Contract {\it No} 06DA453.

\end{document}